\documentstyle[sprocl,epsf]{article}

\catcode`@=11
\def\citenum#1{{\def\@cite##1##2{##1}\cite{#1}}}
\catcode`@=12

\def\eps{\epsilon}
\newcommand{\alt}{\mathrel{\raisebox{-.6ex}{$\stackrel{\textstyle<}{\sim}$}}}
\newcommand{\agt}{\mathrel{\raisebox{-.6ex}{$\stackrel{\textstyle>}{\sim}$}}}

\begin{document}

\vspace*{-.75in}

\font\fortssbx=cmssbx10 scaled \magstep1
\hbox to \hsize{
\hbox{\fortssbx University of Wisconsin - Madison}
\hfill$\vtop{\normalsize\hbox{\bf MADPH-98-1068}
                \hbox{July 1998}}$ }

\vspace{.2in}

\title{\uppercase{Four-Neutrino Oscillations}\footnote{Talk presented at the {\it 6th International Symposium on Particles, Strings and Cosmology (PASCOS\,98)}, Northeastern University, Boston, March 1998.}}
\author{\unskip\smallskip V. BARGER}
\address{Physics Department, University of Wisconsin, Madison, WI 53706}

\maketitle
\thispagestyle{empty}
\abstracts{Models with three active neutrinos and one sterile neutrino can naturally account for maximal oscillations of atmospheric neutrinos, explain the solar neutrino deficit, and accommodate the results of the LSND experiment. The models predict either $\nu_e\to\nu_\tau$ or $\nu_e\to\nu_s$ oscillations in long-baseline experiments with the atmospheric $\delta m^2$ scale and amplitude determined by the LSND oscillations.}

\section{Introduction}

When neutrino flavor eigenstates $\nu_f$ are not the same as the mass eigenstates $\nu_i$, e.g., for two neutrinos, 
\begin{equation}
\nu_f = \cos\theta\nu_1 + \sin\theta\nu_2\,,\qquad
\nu_{f'} = -\sin\theta\nu_1 + \cos\theta\nu_2\,,
\end{equation}
then neutrinos oscillate. The vacuum oscillation probabilities are
\begin{eqnarray}
P(\nu_f\to\nu_{f'}, L) &=& A\sin^2\left(\delta m^2 L
\over 4E\right)\quad\qquad\mbox{``appearance''}\,,\\
P(\nu_f\to\nu_f, L) &=& 1 - A \sin^2 \left(\delta m^2 L
\over 4E\right)\qquad\mbox{``survival''}\,,
\end{eqnarray}
where $A=\sin^22\theta$, $\delta m^2 = m_2^2 - m_1^2$; $L$ is the path length and $E$ is the neutrino energy. 

There is mounting experimental evidence for neutrino oscillations\cite{summaries} from the atmospheric neutrino anomaly (vacuum oscillations), the solar neutrino deficit (matter or vacuum oscillations), and the LSND experiment (vacuum oscillations). Each can be explained by oscillations of two flavors. However, three independent $\delta m^2$ are required, but there are only two independent $\delta m^2$ from $\nu_e$, $\nu_\mu$, and $\nu_\tau$.  If all observed oscillation effects are real, a way out is oscillations to both active and sterile neutrino flavors.\cite{bllp,ourpapers}

Sterile neutrinos have no electroweak interactions (e.g., $Z\not\to \nu_s\nu_s$) and thus evade accelerator constraints. However, $\rm active \leftrightarrow sterile$ oscillation in the early universe would lead to $N_\nu = 4 $ neutrino species by the time of big bang nucleosynthesis (BBN), which is inconsistent with an $N_\nu\alt 2.6$ bound based on low deuterium abundance but allowed by conservative estimates that $N_\nu < 4.5$. If there exists a lepton number asymmetry $L_\nu = (n_\nu - n_{\bar \nu})/n_\gamma > 7.5 \times 10^{-5}$ at the epoch with temperature $T< 10$--20~MeV, then the appearance of sterile neutrinos in the early universe can be suppressed.\cite{foot} In the following, both tightly constrained ($\delta m_{fs}^2 A_{fs} < 10^{-7}\rm\,eV^2$) and unconstrained ($A_{fs}\sim 1$) $\rm active \leftrightarrow sterile$ oscillation possibilities are considered.

\section{The Data}

\underline{\it LSND} \ 
The Los Alamos experiment studied $\bar\nu_\mu\to\bar\nu_e$ oscillations from $\bar\nu_\mu$ of $\mu^+$ decay at rest and $\nu_\mu\to\nu_e$ from $\nu_\mu$ of $\pi^+$ decay in flight. The results, including restrictions from BNL, KARMEN and Bugey experiments, suggest $\nu_\mu\to \nu_e$ oscillations with
\begin{equation}
0.3{\rm\ eV^2} < \delta m_{\rm LSND}^2 <  2.0{\rm\ eV^2}\,,\qquad
A_{\rm LSND} \approx 4\times10^{-2} \ \rm to \ 3\times10^{-3} \,.
\end{equation}

\noindent
\underline{\it Atmospheric} \ 
Cosmic ray interactions with the atmosphere produce $\pi$-mesons and the decays $\pi\to\mu\nu$ and $\mu\to\nu_e e\nu_\mu$ give $\nu_\mu$ and $\nu_e$ fluxes in the approximate ratio $(\nu_\mu+\bar\nu_\mu)/(\nu_e+\bar\nu_e)\sim2$. Measurements of $R=(N_\mu/N_e)_{\rm data} / (N_\mu/N_e)_{\rm MC}$ for $E_\nu\sim 1$~GeV find values of $R\sim 0.6$. In the water Cherenkov experiments the single rings due to muons are fairly clean and sharp, while those from electrons are fuzzy to do electromagnetic showers. The Super-Kamiokande measurements\cite{summaries,kearns} of $R$ versus the zenith angle $\theta$ are shown in Fig.~1a for sub-GeV and multi-GeV energies. As suggested long ago,\cite{oldatmos} the data are well described by $\nu_\mu\to\nu_\tau$ or $\nu_\mu\to \nu_s$ oscillations with $\delta m_{\rm ATM}^2 \approx 5\times10^{-3}\rm\,eV^2$ and $A_{\rm ATM}\approx1$, as shown by the dotted histograms in Fig.~1a. The relation of the path length $L$ to the zenith angle $\theta$ is displayed in Fig.~1b. For sub-GeV neutrino energies, $L/E$ is large at $\cos\theta<0$ and the oscillations average, $P(\nu_\mu\to\nu_\mu)\approx 0.5$. At multi-GeV energies, $L/E$ is large at $\cos\theta=-1$ and $P(\nu_\mu\to\nu_\mu)\approx 0.5$; also $L/E$ is small at $\cos\theta=+1$ and $P(\nu_\mu\to\nu_\mu)\approx 1$. The separate distributions of $\mu$-like and $e$-like events versus the zenith angle establish that the anomalous $R$-ratio is due to a deficit of upward $\mu$-like events. The allowed ranges of $\nu_\mu\to\nu_\tau$ oscillation parameters are summarized in Fig.~2.

\noindent
\underline{\it Solar} \
Three types of solar $\nu_e$ experiments, (i)~$\nu_e$ capture in Cl [Homestake], (ii)~$\nu_e e\to\nu_e e$ [Kamiokande and SuperKamiokande], (iii)~$\nu_e$ capture in Ga, measure rates below standard model expectations. The different experiments are sensitive to different ranges of solar $E_\nu$. There are three regions of oscillation parameter space that can accommodate all these observations:\cite{hata}

\begin{center}
\begin{tabular}{lcc}
& $\delta m^2_{\rm SOL}\rm\ (eV)^2$& $A_{\rm SOL}$\\
Small Angle Matter (SAM)& $\sim 10^{-5}$& $\sim10^{-2}$\\
Large Angle Matter (LAM)& $\sim 10^{-5}$& $\sim 0.6$\ \ \ \\
Vacuum Long Wavelength& $\sim 10^{-1\rlap{\scriptsize0}}$& $\sim1$\ \ \  \ \
\end{tabular}
\end{center}

\noindent
Figure 3 illustrates these parameter regions for the solar solutions along with the regions for the atmospheric and LSND oscillation interpretations. The solar $\nu$ oscillation solutions will eventually be distinguished by use of all the following measurements: (i)~time-averaged total flux, (ii)~day-night dependence (earth-matter effects), (iii)~energy spectra (electron energy in $\nu e \to \nu e$ events), (iv)~seasonal variation, and (v)~the neutral-current to charged-current event ratio (SNO experiment).

\begin{figure}[t]
\centering\leavevmode
\epsfxsize=2.6in\epsffile{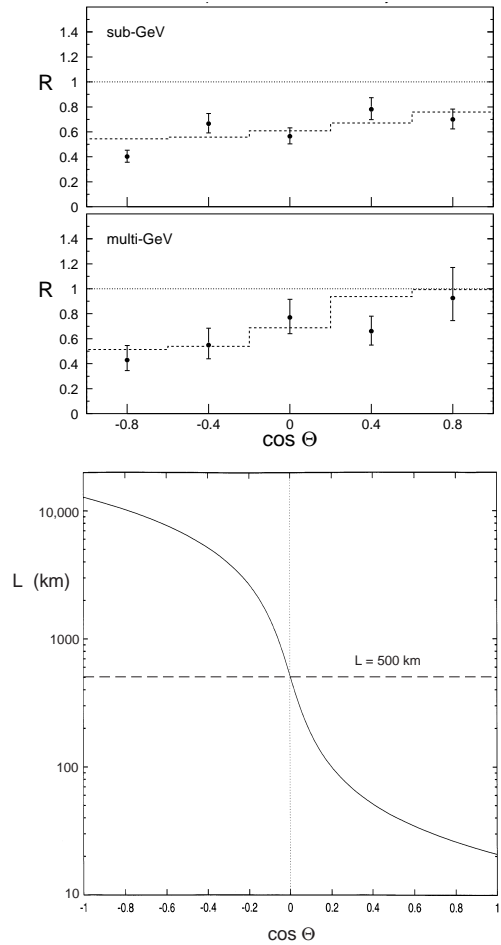}

\caption[]{(a)~The zenith angle dependence of $R$ for sub-GeV and multi-GeV atmospheric neutrino samples from Super-Kamiokande; from Ref.~\citenum{kearns}. (b)~Path length versus zenith angle.}
\end{figure}

\begin{figure}[t]
\centering\leavevmode
\epsfxsize=2.8in\epsffile{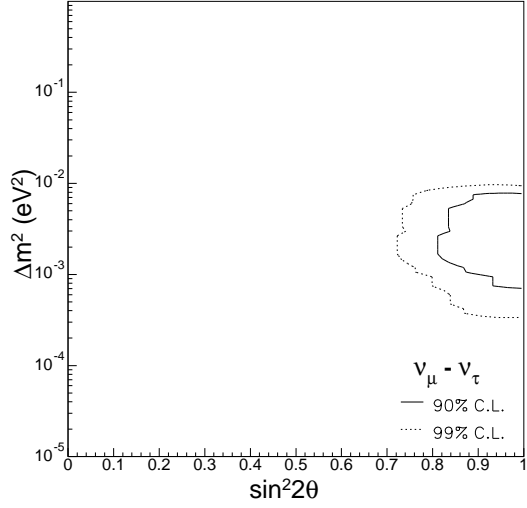}

\caption[]{Confidence intervals for $\sin^22\theta$ and $\Delta m^2$ based on a $\chi^2$ fit to Super-Kamiokande atmospheric neutrino data. The solid line is 90\% CL, the dashed line is 99\% CL. From Ref.~\citenum{kearns}.}
\end{figure}

\begin{figure}[b]
\centering\leavevmode
\epsfxsize=2.3in\epsffile{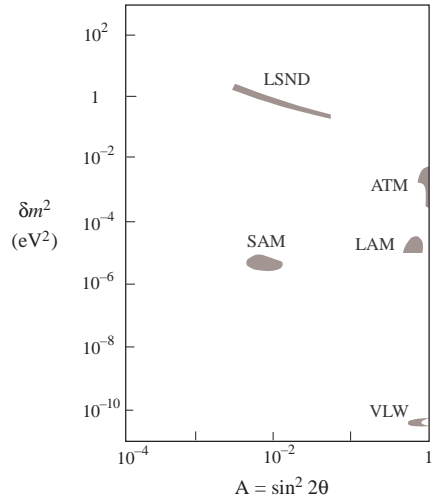}

\caption[]{The three allowed two-neutrino solar solutions for
$\nu_e\rightarrow\nu_\tau$ oscillations. 
The corresponding regions for $\nu_e\rightarrow\nu_s$ oscillations are
similar to the $\nu_e\rightarrow\nu_\tau$ case.}
\end{figure}

\clearpage

\section{4-Neutrino Models}

Table 1 shows the options for oscillation solutions to all data. The preferred mass spectrum is two nearly degenerate mass pairs separated by the LSND scale, as displayed in Fig.~4. Here $m_0 > m_1$ for the case of matter oscillations so that $\nu_e$ is resonant\cite{wolf} in the Sun. The alternative of a $1+3$ mass hierarchy with one heavier mass scale separated from three lighter, nearly degenerate states is disfavored when the null results of reactor and accelerator disappearance experiments are taken into account.\cite{bilenky}

\begin{center}
\leavevmode
\epsfxsize=2.5in\epsffile{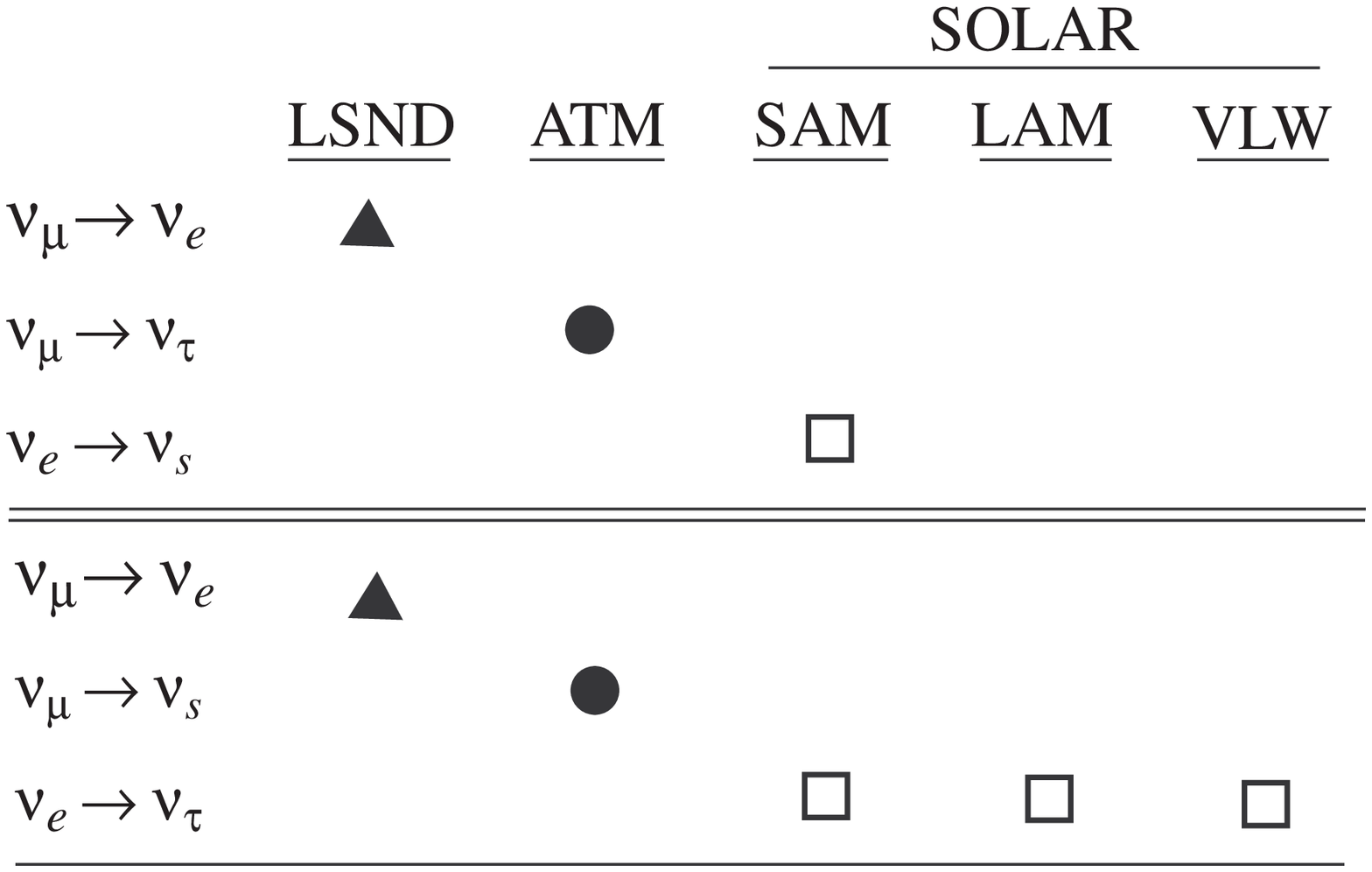}

\smallskip
\footnotesize Table 1: Four-neutrino oscillation possibilities.
\end{center}

\bigskip

\begin{figure}[h]
\centering\leavevmode
\epsfxsize=1.7in\epsffile{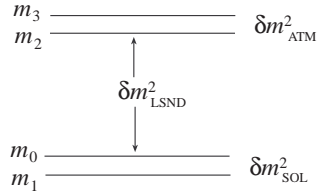}

\caption{Neutrino mass spectrum showing which mass splittings are responsible for the LSND, atmospheric, and solar oscillations.}
\end{figure}

Consider first the case with $\nu_\mu\to\nu_\mu$ for the atmospheric oscillations. A~neutrino mass matrix of the form\cite{ourpapers,gibbons}
\begin{equation}
{ \hspace{1.5em}\begin{array}{cccc} \nu_s& \nu_e& \nu_\mu& \nu_\tau \end{array} \atop
M = m \left(
\begin{array}{cccc}
\eps_1& \eps_2& 0 & 0 \\
\eps_2& 0 & 0 & \eps_3\\
0& 0& \eps_4& 1\\
0& \eps_3& 1& \eps_4
\end{array} \right)
\begin{array}{c} \nu_s\\ \nu_e\\ \nu_\mu\\ \nu_\tau \end{array} }
\end{equation}
can reproduce the three observed $\delta m^2$, the amplitudes $A_{\rm LSND}$ and $A_{\rm SOL}$, and naturally give $A_{\rm ATM}=1$. Here the $\epsilon_i$ are all small compared to unity. The values of $\eps_1$ and $\eps_2$ determine which of the three solar solutions is realized. The mixing matrix corresponding to this mass matrix is
\begin{equation} \def\arraystretch{1.4}
\left(\begin{array}{c}\nu_s\\ \nu_e\\ \nu_\mu\\ \nu_\tau \end{array} \right) = 
\left( \begin{array}{cccc}
\cos\theta& -\sin\theta& 0& 0\\
\sin\theta& \phantom-\cos\theta& {1\over\sqrt 2} \eps_3& {1\over \sqrt 2}\eps_3\\
-\eps_3\sin\theta& -\eps_3\cos\theta& {1\over\sqrt2}& {1\over\sqrt 2}\\
0& 0& \llap{$-$}{1\over\sqrt2}& {1\over\sqrt2}
\end{array}\right)
\left(\begin{array}{c}\nu_0\\ \nu_1\\ \nu_2\\ \nu_3\end{array} \right)
\end{equation}
The vacuum probabilities in this model are
\begin{eqnarray}
\rm LSND && P(\nu_\mu\leftrightarrow\nu_e) = 4\eps_3^2 \sin^2 \left(m^2L\over 4E\right)    \\
\rm ATM  && P(\nu_\mu\leftrightarrow\nu_\tau) \simeq \sin^2 \left(m^2\eps_4L\over E\right) \\
\rm SOLAR && P(\nu_e\leftrightarrow\nu_s) \simeq {4\eps_2^2\over \eps_1^2+4\eps_2^2} \sin^2 \Delta_{\rm SOL}   
\end{eqnarray}
where $\Delta_{\rm SOL} = \delta m^2_{\rm SOL} L/E$, $\delta m_{\rm SOL}^2 \simeq 4\eps_2^2/\eps_1^2$, $4\eps_2^2/(\eps_1^2+\eps_2^2)$, and 1, for SAM, LAM, and VLW, respectively.
Matter effects must be included in the solar SAM and LAM solutions.

By construction this model has effective two-neutrino oscillation solutions for LSND, ATM, and solar phenomena. The model makes a number of predictions:\cite{ourpapers}

(i) Neutrinoless double-$\beta$ decay vanishes at tree level because $M_{\nu_e\nu_e} = 0$.

(ii) The neutrino mass spectrum is $m_3, m_4\simeq 1.4$~eV, $m_0 \simeq 2\times10^{-3}$~eV, $m_1\simeq 4\times10^{-6}$~eV. There will be no measurable effect at the endpoint of tritium beta decay if $\nu_e$ is primarily associated with the lighter pair.

(iii) The hot dark matter contribution to the mass density of the Universe is $\Omega_\nu = \sum (m_\nu/93{\rm\ eV}) h^{-2}$, where $h\simeq0.65$. The SLOAN Digital Sky Survey is expected to have sensitivity down to $m_\nu = 0.2$--0.9~eV for two nearly degenerate neutrinos,\cite{hu} which covers the interesting range from LSND.

(iv) In the SNO solar experiment, both CC and NC event rates would be suppressed, with $\rm NC/CC = 1$, if $\nu_e\to\nu_s$ is the solar solution.

(v) In reactor experiments the $\bar\nu_e$ disappearance $P(\bar\nu_e\to\bar\nu_e) =1 - \break  A_{\rm LSND}\sin^2\Delta_{\rm LSND}$ with $A_{\rm LSND} \approx 2.5\times10^{-3}$ is not detectable. For example, the CHOOZ experiment sensitivity is $A\agt 0.2$ for $\delta m^2 \agt 10^{-3}\rm\,eV^2$.

(vi) Long-baseline experiments with $L/E\approx 10$--$10^2$~km/GeV could measure $P(\nu_\mu\leftrightarrow\nu_\tau) \simeq \sin^2\Delta_{\rm ATM}$ and confirm the atmospheric oscillation result. In addition, the prediction of new oscillations 
\begin{eqnarray}
P(\nu_\mu\leftrightarrow\nu_e) &\simeq& a(2-\sin^2\Delta_{\rm ATM})\\
P(\nu_e\leftrightarrow\nu_\tau) &\simeq& a\sin^2\Delta_{\rm ATM}
\end{eqnarray}
with $a = A_{\rm LSND}/4 \approx 10^{-2}$ to $10^{-3}$ could be tested. For this purpose intense neutrino beams are required. The MINOS experiment (Fermilab to Soudan) could confirm the $\nu_\mu\to\nu_\tau$ oscillations and test the $\nu_\mu\to\nu_e$ prediction, provided that $\delta m_{\rm ATM}^2 > 2\times10^{-3}\rm\,eV^2$.

In the future, a special purpose muon storage ring could provide high intensity neutrino beams with well-determined fluxes that could be directed towards any detector on the earth.\cite{geer} It could be possible to store $\sim10^{21}$ $\mu^+$ or $\mu^-$ per year and obtain $\sim10^{20}$ neutrinos from the muon decays. Oscillations give ``wrong sign'' leptons from those produced by the beam. For example, $\mu^-$ decays give $\bar\nu_e$ and $\nu_\mu$ fluxes so detection of $\mu^+, e^-, \tau^\pm$ leptons tests for $\bar\nu_e\to\bar\nu_\mu(\bar\nu_\tau)$ and $\nu_\mu\to\nu_e(\nu_\tau)$ oscillations. Taus can be detected via their $\tau\to\mu$ decays and the $\tau$-charges so determined to distinguish $\nu_\mu\to\nu_\tau$ and $\bar\nu_e\to\bar\nu_\tau$ oscillations. The ranges of oscillation parameters that could be tested in such long-baseline experiments is illustrated in Fig.~5. 

\begin{figure}[t]
\centering\leavevmode
\epsfxsize=2.5in\epsffile{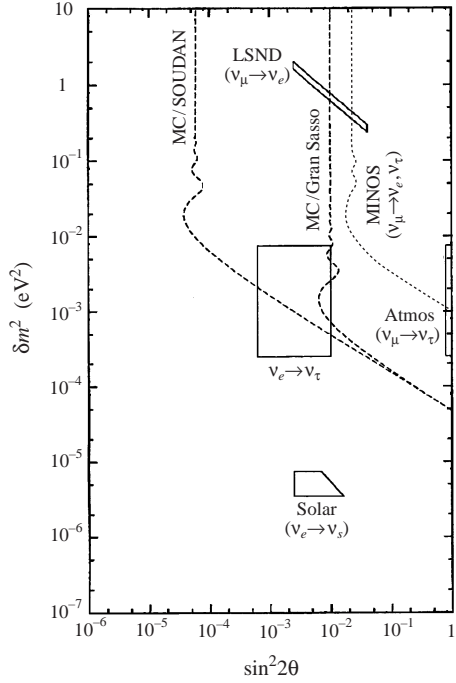}

\caption[]{Predicted region in the effective $\delta
m^2$-$\sin^22\theta$ parameter space for $\nu_e \rightarrow \nu_\tau$
oscillations in the four-neutrino model (solid rectangle), which is
determined by ${1\over4}$ of the
LSND $\nu_\mu \rightarrow \nu_e$ oscillation amplitude and the
atmospheric neutrino $\nu_\mu \rightarrow \nu_\tau$ oscillation
$\delta m^2$ scale. The dotted curves show the potential limits on
$\nu_\mu\rightarrow \nu_e, \nu_\tau$ oscillations from the MINOS
experiment 
and the dashed curves show the potential limits
on $\nu_e,\nu_\mu\rightarrow\nu_\tau$ oscillations that can be set by
neutrino beams from an intense muon source at
Fermilab 
to detectors at the SOUDAN and GRAN SASSO sites for
muons with energy of 20~GeV. Also shown are the parameters for the solar
$\nu_e \rightarrow \nu_s$ small-angle MSW oscillation.}
\end{figure}

\section{Alternative 4-Neutrino Mixings}

The preceding model assumed large $\nu_\mu\leftrightarrow\nu_\tau$ mixing as the explanation of the atmospheric data. The alternative scenario with large $\nu_\mu\leftrightarrow\nu_s$ mixing is obtained in a straightforward manner by interchange of $\nu_s$ and $\nu_\tau$ labels. Then the long-baseline predictions are $\nu_\mu$ disappearance
\begin{equation}
P(\nu_\mu\leftrightarrow\nu_s) \simeq \sin^2\Delta_{\rm ATM}
\end{equation}
and $\nu_\mu\leftrightarrow\nu_e$ appearance
\begin{equation}
P(\nu_\mu\leftrightarrow\nu_e) \simeq a(2-\sin^2 \Delta_{\rm ATM}) \,,
\end{equation}
A more general scenario could have large $\nu_\mu$ mixing with a linear combination of $\nu_s$ and $\nu_\tau$.

\section{Summary}

A simple mass matrix for four neutrinos with strategically placed zeros can accommodate the indications for neutrino oscillations from the LSND, atmospheric, and solar data. The two principle variants of the model have oscillations of
\begin{eqnarray}
 \nu_\mu\to\nu_\tau\mbox{ or }\nu_s  &&\rm ATM \\
 \nu_e\to\nu_s\mbox{ or } \nu_\tau  &&\rm SOLAR 
\end{eqnarray}
with $\nu_\mu\to\nu_e$ for LSND. The predictions for long-baseline experiments are oscillations with the $\delta m^2_{\rm ATM}$ scale with maximal amplitude in the channel
$\nu_\mu\to\nu_\tau\mbox{ or }\nu_s$
and amplitudes $\approx 10^{-2}$ to $10^{-3}$ in the channels
$\nu_e\to\nu_\tau\mbox{ or }\nu_s$
and $ \nu_\mu\to\nu_e$.

\section*{Acknowledgments}
\unskip\smallskip{\small
I thank Sandip Pakvasa, Tom Weiler and Kerry Whisnant for collaborations on neutrino oscillation analyses and for their comments in the preparation of this report.
This research was supported in part by the U.S.~Department of Energy under Grant No.~DE-FG02-95ER40896 and in part by the University of Wisconsin Research Committee with funds granted by the Wisconsin Alumni Research Foundation.}

\vspace{-0.1in}

\section*{References}
\unskip

\smallskip
\begin{thebibliography}{99}
\frenchspacing
\baselineskip=11pt

\bibitem{summaries}
For recent summaries of the data see e.g., {\it Proceedings of the ITP Conference on Solar Neutrinos}, Santa Barbara, Dec.~1997 and {\it Proceedings of Neutrino 98}, Takayama, Japan, June~1998.

\bibitem{bllp}
V.~Barger, P.~Langacker, J.~Leveille, and S.~Pakvasa, Phys. Revl. Lett. {\bf 45}, 962 (1980); D.O.~Caldwell and R.N.~Mohapatra, Phys. Rev. {\bf D48}, 3259 (1993); J.T.~Peltoniemi and J.W.F.~Valle, Nucl. Phys. {\bf B406}, 409 (1993); J.J.~Gomez-Cadenas and M.C.~Gonzalez-Garcia, Z.~Phys. {\bf C71}, 443 (1996); S.M.~Bilenky, C.~Guinti, and W.~Grimus, hep-ph/9711416; E.J.~Chun, A.S.~Joshipura, and A.Y.~Smirnov, Phys. Rev. {\bf D54}, 4654 (1996); K.~Benakli and A.Y.~Smirnov, Phys. Rev. Lett. {\bf 79}, 4314 (1997); Q.Y.~Liu and A.Y~Smirnov, Nucl. Phys. {\bf B524}, 505 (1998).

\bibitem{ourpapers}
This report is largely based on papers by V.~Barger, T.J.~Weiler, and K.~Whisnant, Phys. Lett. {\bf B427}, 97 (1998), and  V.~Barger, S.~Pakvasa, T.J.~Weiler, and K.~Whisnant, hep-ph/9806328. More extensive references to the literature can be found therein.

\bibitem{foot}
R. Foot and R.R. Volkas, Phys. Rev. {\bf D55}, 5147 (1997).

\bibitem{kearns}
E. Kearns,  hep-ex/9803007, to be published in the proceedings of {\it TAUP\,97}, Gran Sasso, Italy, Sept.~1997. More recent Superkamikonda results can be found in Y.~Fukuda et al. (the Super-Kamiokande Collaboration), hep-ex/9807003.

\bibitem{oldatmos}
J.G. Learned, S. Pakvasa, and T.J. Weiler, Phys. Lett. {\bf B207}, 79 (1988);
V. Barger and K. Whisnant, Phys. Lett. {\bf B209}, 365 (1988);
K. Hidaka, M. Honda, and S. Midorikawa, Phys. Rev. Lett. {\bf 61}, 1537
(1988).

\bibitem{hata}
See e.g., N. Hata and P. Langacker, Phys. Rev. {\bf D56}, 6107 (1997);
 a recent update of solar oscillation fits can be found in J.~Bahcall, P.~Krastev, and A.~Smirnov,  hep-ph/9807216.


\bibitem{wolf}
L.~Wolfenstein, Phys. Rev. {\bf D17}, 2369 (1978); S.P.~Mikheyev and A.~Smirnov, Yad. Fiz. {\bf 42}, 1441 (1985); Nuovo Cim. {\bf 9C}, 17 (1986); V.~Barger, N.~Deshpande, P.B.~Pal, R.J.N.~Phillips, and K.~Whisnant, Phys. Rev. {\bf D43}, 1759 (1991).

\bibitem{bilenky}
S.M.~Bilenky, C.~Guinti, and W.~Grimus, Eur. Phys. J. {\bf C1}, 247 (1998).

\bibitem{gibbons}
A somewhat similar mass matrix is considered by S.C.~Gibbons, R.N.~Mohapatra, S.~Nandi, and A.~Raychaudhuri, hep-ph/9803299.

\bibitem{hu}
W.~Hu, D.~Eisenstein, and M.~Tegmark, Phys.\,Rev.\,Lett.\,{\bf 80}, 5255 (1998).

\bibitem{geer}
S. Geer, Phys. Rev. {\bf D57}, 6989 (1998).

\end{thebibliography}
\end{document}